\documentclass[pre,aps]{revtex4}

\usepackage{graphicx}%
\usepackage{amsmath}

\begin{document}
\title{Bessel beam propagation: Energy localization and velocity}
\author{D. Mugnai\footnote{E-mail: d.mugnai@ifac.cnr.it} and I. Mochi\footnote{Permanent address:
Osservatorio Astrofisico di Arcetri, L.go E. Fermi 5, 50125
Firenze, Italy}} \affiliation{``Nello Carrara'' Institute of
Applied Physics-CNR,
\\ Via Panciatichi 64, 50127 Firenze, Italy}

\begin{abstract}
\vspace{1 cm}

The propagation of a Bessel beam (or Bessel-X wave) is analyzed on
the basis of a vectorial treatment. The electric and magnetic
fields are obtained by considering a realistic situation able to
generate that kind of scalar field. Specifically, we analyze the
field due to a ring-shaped aperture over a metallic screen on
which a linearly polarized plane wave impinges. On this basis, and
in the far field approximation, we can obtain information about
the propagation of energy flux and the velocity of the energy.
\end{abstract}

\maketitle

\newcommand{\be}{\begin{equation}}
\newcommand{\ee}{\end{equation}}
\newcommand{\bea}{\begin{eqnarray}}
\newcommand{\eea}{\end{eqnarray}}

The motion of a Bessel beam  is of great interest in physics both
for its characteristic as a non-diffracting beam
\cite{dur87,dur87-1,zio,spr,dur91,tan,rei}, and for its
implications with regard to the topic of superluminality
\cite{saa,mug00,ale,bes,zam,saa1}.
  Extended studies have been devoted to these subjects from both an experimental and a theoretical
point of view. However, in spite of the many efforts devoted to
this topic, no definite answer has been found about the amount of
the energy transfer and its velocity.

\begin{figure}[t]
\includegraphics[width=0.6\textwidth]{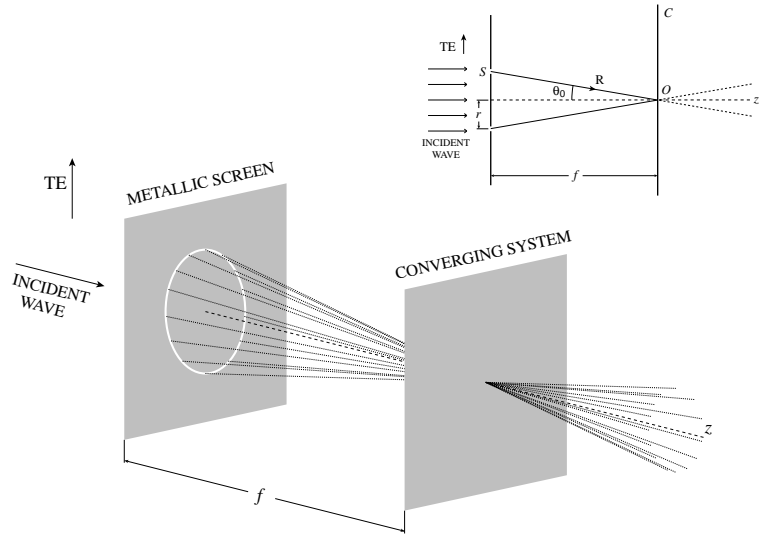}
 \caption{Scheme of the system considered for the theoretical analysis presented here.
The metallic screen (over which the ring-shaped aperture is
placed) and the converging system must be considered as being of
infinite dimension. Details are shown in the inset.}
\label{scheme}
\end{figure}

As far as Bessel beam propagation is concerned, the problem is
mainly related to the difficulty in finding the vectorial field
that describes this system, while, on the contrary,
 the field in the scalar approximation is well-known.

The purpose of the present work is to investigate the propagation
of a Bessel beam on the basis of a vectorial treatment. This kind
of approach allows us to obtain information regarding the
propagation of the energy flux and the energy mean velocity.

A Bessel beam  consists of a set of plane waves  with directions
of propagation {\bf s} = $\alpha_1${\bf i}+ $\beta_1${\bf j} +
$\gamma_1${\bf k} which makes the same angle $\theta_0 \, (0 \leq
\theta_0 < \pi /2$) with the $z$-axis (hence
 $\gamma_1 =\cos\theta_0$ for all the plane waves).
 In spherical coordinates ($\rho,\,\theta,\,\varphi$), the direction of propagation is specified by

\bea
 \alpha_1 =\sin\theta_0\cos\varphi\,,\:\:\:\beta_1
=\sin\theta_0\sin\varphi\, \:\:\: \gamma_1 =\cos\theta_0 \,.
 \eea
Thus, for  propagation in vacuum or air, each one of these waves,
at the point $x,\:y,\:z,$ can be written as

\be
 u (P)=u_0 d\varphi \, \exp [ik_0(\alpha_1 x+\beta_1 y+\gamma_1 z)]
\, \exp(-i\omega t),
  \label{onda}
 \ee
  where $u_0 d\varphi$ is the amplitude of the elementary wave, $\omega$ is the
angular frequency, $k_0=\omega /c$ is the wavenumber, and $x, \:y$
and $z$ denotes the Cartesian coordinates of $P$. In cylindrical
coordinates $\rho,\,\psi,\, z$ around the $z$-axis

\bea
 x= \rho\cos\psi,\:\:\: y= \rho\sin\psi, \:\:\:  z\equiv z,
\nonumber \eea
 and the  total field $U$,  given by the superposition
of all the waves (\ref{onda}), can be obtained by integrating over
$d\varphi$, that is,

\bea
 U = &u_0&\int_0^{2\pi}\,\exp [ik_0 (\alpha_1 x+\beta_1
y+\gamma_1 z)]\,\exp(-i\omega t)\,d\varphi \nonumber \\
 =&& 2\pi
u_0 J_0(k_0\rho\sin\theta_0 )\exp \left(i\omega \frac{z}{c}\,
\cos\theta_0 \right)\,\exp(-i\omega t)
 \label{bes}
 \eea
where $J_0$ denotes the zero-order Bessel function of first
kind\cite{wat}.

The scalar field of Eq. (\ref{bes}) is known as a Bessel beam (or
Bessel-X wave), the unusual features of which are
\newline
- that it does not change its shape during propagation, since its
amplitude is independent of $z$\cite{note};
\newline
- that it propagates in the $z$ direction with phase and group
velocities $v=c/\cos\theta_0$  larger than
$c$\cite{saa,mug00,ale,note1}.
\newline
  Both the above mentioned characteristics can be analyzed in detail
by means of a vectorial treatment, since the  scalar field
(\ref{bes}) represents an approximation of an electromagnetic
field, and only a knowledge of the vectorial field (and of the
Poynting vector in particular) can provide detailed information
about the energy propagation.

Vectorial fields with amplitude proportional to the zero-order
Bessel function can be found in different ways. However, in order
to derive just the vectorial field describing a system which has
Eq. (\ref{bes}) as scalar approximation, we have to consider a
realistic situation that is able to generate a field of that kind.
For this purpose, let us consider the system of Fig. \ref{scheme},
which consists of a ring-shaped aperture, of radius $r$, over a
metallic screen on which a linearly polarized plane wave impinges
(from the left). The ring is placed on the focal plane of the
converging system $C$ with focal length $f\gg \lambda$, $\lambda$
being the wavelength. Let us consider the impinging electric filed
to be polarized in the {\bf i} direction, and the thickness $d$ of
the ring to be very small with respect to $\lambda$.
 We may assume the element $dA$ of the ring, at $S(r,\varphi,-f)$, to
behave as an elementary dipole, parallel to {\bf i}, with
amplitude proportional to $d\varphi$ (this requires a suitable
choice of the transparency of the ring at $S$).
 Thus, the associated field at the optical center $O$ of $C$ has the well-known
characteristics of the far field radiated by an elementary dipole,
that is to be a spherical wave centered at $S$, with the electric
field {\bf e} in the meridional plane of the dipole through $O$,
and perpendicular to the direction {\bf R} from $S$ to $O$.
 We can write

\bea &{\bf R}&= -r\cos\varphi\,{\bf i}
 - r\sin\varphi \,{\bf j}
  +f \,{\bf k}, \nonumber \\
  &{\bf e}&=e_x{\bf i}+e_y{\bf j}+e_z{\bf k}.
 \eea
The two characteristics mentioned above can be written as

\bea
 {\bf e}\cdot{\bf R}=-r\cos\varphi \,e_x-r\sin\varphi \,e_y+f\,e_z=0 \nonumber \\
({\bf R}\times{\bf e})\cdot {\bf i}=f \,e_y+r\sin\varphi \,e_z=0,
\label{sys}
 \eea
 where we disregarded a phase factor $\exp (ik R)$.
By solving Eqs. (\ref{sys}) we obtain

\bea
 e_y=-r\cos\varphi \, e_x\left(\frac{r\sin\varphi}{r^2\sin^2\varphi+f^2}\right)
  \label{ey} \\
 e_z=r\cos\varphi \, e_x\left(\frac{f}{r^2\sin^2\varphi+f^2}\right).
 \label{ez}
\eea
  At this point, we can assume that the field emerging from
the optical converging system is a plane wave propagating in the
direction of {\bf R}, with amplitude proportional to the amplitude
at $C$ of the incident field, namely \cite{note2}

\bea
 {\bf e}&=& \left(e_x{\bf i}+e_y{\bf j}+e_z{\bf k}\right)\exp
[i k(\alpha x+\beta y +\gamma z)] \label{e}  \\
  {\bf h}&=& \frac{1}{Z} \left[\frac{}{}(\alpha{\bf i}+\beta{\bf j}+\gamma{\bf
k})
   \times \left(e_x{\bf i}+e_y{\bf j}+e_z{\bf k}\right)\right]\exp [i k(\alpha x+\beta y +\gamma
   z)]
\label{h} \eea where $\alpha =-\cos\varphi \sin\theta_0,\:\beta =-
\sin\varphi \sin\theta_0,\:\gamma =\cos\theta_0$ are the director
cosines, $Z$ is the free-space impedance, and the temporal factor
$\exp (-i\omega t)$ is omitted. The total electric field ${\bf E}$
will be given by the superposition of
 all {\bf e}$d\varphi$ contribution arising from the dipoles, and results in

\bea
 {\bf E} &=& \int_0^{2\pi} {\bf e}\:d\varphi = \int_0^{2\pi}
(e_x{\bf i}+e_y{\bf j}+e_z{\bf k}) \exp \left\{i
k\left[-\rho\left(\frac{r}{R}\right)\cos(\varphi -\psi )+
\cos\theta_0 z\right]\right\} \:d\varphi \nonumber \\
  &=& \exp \left( i k \cos\theta_0 z \right)
 \int_0^{2\pi} (e_x{\bf i}+e_y{\bf j}+e_z{\bf k})
\exp \left [-i k \rho \sin\theta_0\cos(\varphi -\psi )\right]
d\varphi  .
 \label{campotot}
 \eea
 With reference to Eqs. (\ref{ey}) and (\ref{ez}), it is expedient
 to choose

 \be
 e_x= \frac{e_0}{f^2} \, ( r^2\sin^2\varphi+f^2):
 \label{ex}
 \ee
 this condition can be experimentally obtained
 by a suitable choice of the transparency of the ring as a function of
 $\varphi$.
  Thus, by substituting Eqs. ({\ref{ey}), (\ref{ez}) and (\ref{ex})
  into Eq. (\ref{campotot}), and by recalling that
  $r=f\tan\theta_0$, we finally obtain (calculations are rather cumbersome but
 of no difficulty)

\bea E_x &=& 2\pi\,e_0 e^{ i  \xi z}
 \left\{  J_0(\eta\rho)+ \tan^2 \theta_0
  \left[\left(J_0(\eta\rho)-\frac{J_1(\eta\rho)}{\eta\rho}\right)-
  \cos^2\psi  \left(J_0(\eta\rho)-\frac{2J_1(\eta\rho)}{\eta\rho}\right)\right]\right\}
   \label{ex1} \\
 E_y &=& - 2 \pi \,e_0 e^{ i  \xi z }
  \left[  \, \frac{\sin 2\psi}{2} \tan^2 \theta_0  \left(J_0(\eta\rho)-\frac{2J_1(\eta\rho)}{\eta\rho}\right)\right]
  \label{ey1}  \\
E_z &=& - 2\pi i \,e_0 e^{ i  \xi z }
  \left [ \tan \theta_0 \cos\psi J_1(\eta\rho) \right],
 \label{ez1}
  \eea
where $\xi =k\cos\theta_0, \:\eta =k \sin\theta_0$, and $J_1$
denotes the first-order Bessel function of first kind.

\begin{figure}[t]
\includegraphics[width=0.5\textwidth]{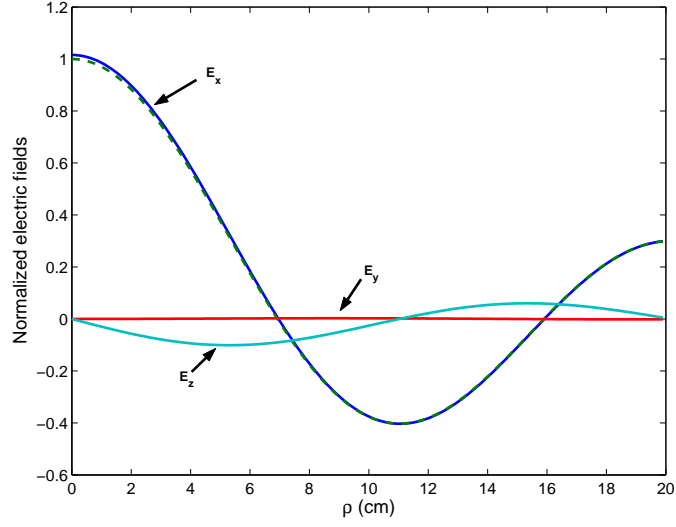}
 \caption{Electric fields $E_x,\:E_y$ and $E_z$ normalized to $2\pi e_0 \exp (i\xi z)$
  (continuous lines) vs. $\rho$, as given by
 Eqs. (\ref{ex1})-(\ref{ez1}), for $k=2,\:\psi =10^\circ$ and
 $\theta_0  =10^\circ$. In $E_z$ phase factor $e^{i\pi /2}$ is disregarded. The dashed line represents
  the normalized scalar field (\ref{bes})  for the same parameter values.}
 \label{field}
\end{figure}

Equation (\ref{ex1}) (the main contribution of the electric field)
describes a field different from the scalar field of Eq.
(\ref{bes}) because of the presence of the term depending on
$\tan\theta_0$. However, for $\theta_0\ll\pi /2\: ( r \ll f )$, as
in the present case, this term is negligible. We note that also
the dependence on $\psi$ (which is absent in the scalar
approximation) is negligible, and may be due to the approximation
indicated in \cite{note2}.
  In Fig. \ref{field}  we
report the normalized value of $E_x,\:E_y$ and $E_z$ vs $\rho$,
for $\theta_0 =10^\circ$, together with the scalar field of Eq.
(\ref{bes}).  The scalar field is practically coincident with
$E_x$. Therefore, we can conclude that the vectorial field derived
above has Eq. (\ref{bes}) as its scalar approximation, at least
for $r\ll f$.

The magnetic field can be derived by Eq. (\ref{h}), and results in

\bea H_x &=&  0
   \label{hx} \\
 H_y &=& \frac{2\pi}{Z}\: e_0 e^{i  \xi z}
   \, \frac{1}{\cos\theta_0}\,  J_0(\eta\rho)
  \label{hy}  \\
H_z &=&
 - i \frac{2\pi}{Z}\: e_0 e^{ i  \xi z} \sin\psi
  \, \frac{\sin\theta_0}{\cos^2\theta_0} \,  J_1(\eta\rho)
   \label{hz}
  \eea
From a knowledge of the electric and magnetic fields, we are now
in a position to evaluate the mean density of the energy flux
which is defined as one half of the real part of the complex
Poynting vector\cite{jac,str}

\begin{figure}[hbt]
\includegraphics[width=0.5\textwidth]{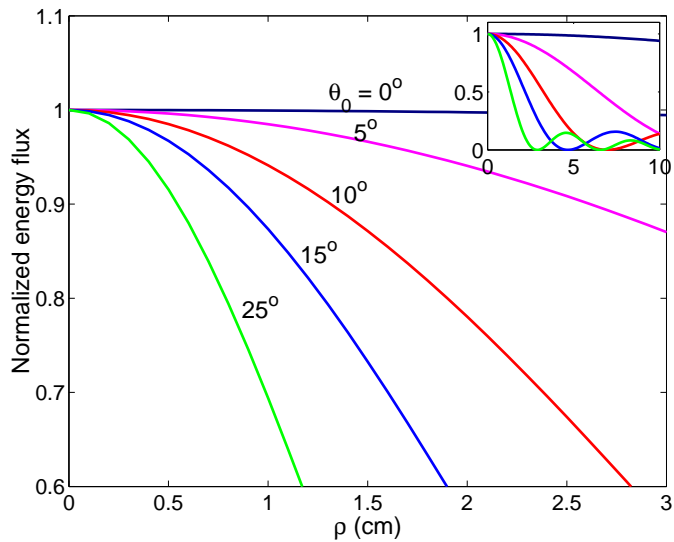}
\caption{Energy flux $S_z$ as given by Eq. (\ref{sz}) normalized
to its maximum value vs $\rho$, for a few values of $\theta_0$.
For the sake of completeness, in the inset $S_z$ is shown for
larger values of $\rho$, even if only the region around the main
peak is of physical interest. Parameter values are as in Fig.
\ref{field}.}
 \label{flux}
\end{figure}

 \be
{\bf S}=\frac{1}{2}\, {\rm Re} \left( {\bf E}\times {\bf
H}^\star\right).  \ee
 For the fields (\ref{ex1})-(\ref{ez1}) and (\ref{hx})-(\ref{hz}),
it turns out  that $S$ has only the $k$-component

\be
 S_z = \frac{1}{2} \left(E_x H_y^\star\right),
 \label{sz}
\ee
 that is the propagation of the energy flux occurs only in the $z$-direction,
in accordance with the information given by the scalar field
(\ref{bes}). Moreover, since the flux is independent of $z$, the
energy propagates with no deformation.

In Fig. \ref{flux}, the behavior of the energy flux (\ref{sz}) is
shown as a function of $\rho$ for a few values of $\theta_0$. We
note that, for $\theta_0 $ very small (nearly plane wave) the flux
is nearly independent of $\rho$, while when the beam originates
the flux increases by increasing $\theta_0$, and tends to
concentrate near $\rho =0$, that is, along the $z$-axis. Thus, for
small values of $\rho$ (that is, in the proximity  of the
$z$-axis), the power supplied by a Bessel beam is always greater
than the one due to a plane wave.

As for the velocity of the energy, from Eq. (\ref{bes}) it follows
that in the scalar approximation the dependence of the field on
$t$ and $z$ occurs only through the quantity $(z/c) \,
\cos\theta_0 -t $ and, therefore, the field propagates with
velocity $ v= c/\cos\theta_0$. On the basis of these arguments, it
could be concluded that also the energy propagates with a velocity
$v_e$ greater than $c$. In the vectorial treatment we can evaluate
the energy velocity $v_e$ as \cite{jac,str}

\begin{figure}[hbt]
\includegraphics[width=0.5\textwidth]{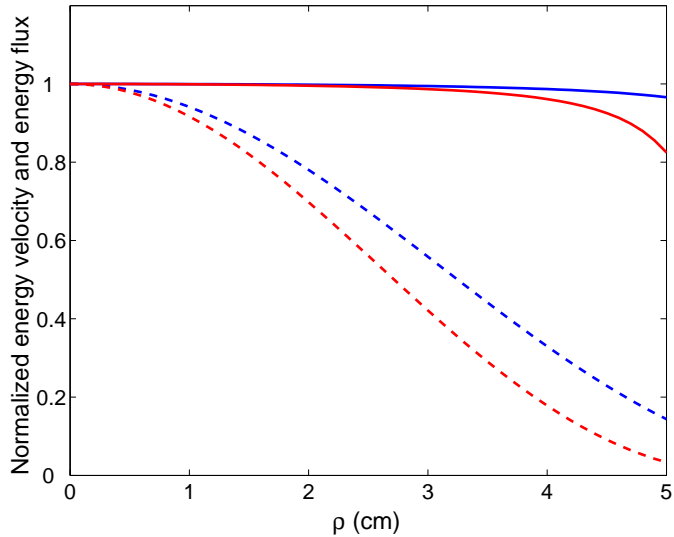}
 \caption{Energy velocity normalized to the light velocity vs
 $\rho$, for $k=2,\:\psi =10^\circ$,
 $\theta_0  =10^\circ$ (blue line) and for  $\theta_0  =12^\circ$
 (red line).
 Dashed lines represent the energy flux relative to the same parameter values.  }
 \label{velocity}
 \end{figure}

\be
 v_e= \frac{S_z}{\frac{1}{4}\left( \varepsilon E\cdot E^\star+\mu H\cdot  H^\star\right)} ,
 \label{v}
  \ee
where the quantity $(1/4)( \varepsilon E\cdot E^\star+\mu H\cdot
H^\star )$ is the total mean density of energy which can be
evaluated with the help of  Eqs. (\ref{ex1})-(\ref{ez1}) and
(\ref{hx})-(\ref{hz}).

In Fig. \ref{velocity}, we report the normalized velocity of
energy as a function of the radial coordinate $\rho$, for
$\theta_0 = 10^\circ$. The velocity is found to be equal to $c$
from $\rho =0$ up to near to the first zero of the Bessel
function: that is, the beam moves like an almost rigid system, in
spite of its dependence on $\rho$ and $\psi$. In the proximity of
the first zero of the Bessel function, the velocity decreases and
tends to zero. Naturally, the zero in the velocity does not
represent a stop of the motion but, more simply, the absence of
energy flux. In this situation, the concept of velocity has no
physical meaning.

Some remarks must be made on this surprising result. In fact, we
recall that for propagation in vacuum ``if an energy density is
associated with the magnitude of the wave $\ldots\ldots$ the
transport of energy occurs with the group velocity, since that is
the rate of which the pulse travel along''\cite{jac7-8}. If the
definition of the energy velocity as given by Eq. (\ref{v}) is
applicable also to a Bessel beam (or, more generally, to localized
waves), it is not clear what kind of physical mechanism makes the
energy velocity different from the phase and group ones.

We wish to recall that the present analysis was performed for an
ideal system in the far field approximation.  For a real system,
we have to take into account the finite dimension of the
converging system, which limits the field depth and introduces
diffractive effects. The role of diffraction, together with the
analysis of the near field (as in real experimental situations),
make the problem much more complicated, and is beyond the purpose
of this paper.

\vspace{0.5 cm}

 {\bf Acknowledgments} \newline
 Special thanks are due to Laura
Ronchi Abbozzo for useful suggestions and discussions.


\begin{thebibliography}{99}
\bibitem{note}  The situation is similar to what
occurs  when only two plane waves interfere, the only difference
being that the two-wave interference pattern occupies the whole
space, while the field (\ref{bes}) is practically limited to a
restricted zone of space, that is, to the first zero of the  $J_0$
function. In this connection, it is worth noting that the field of
Eq. (\ref{bes}) is not properly a beam, since it is not limited by
a caustic surface, inasmuch as $J_0$ oscillates when its argument
tends to infinity.
\bibitem{note1}
 With respect to the monocromatic wave of Eq. (\ref{bes}),
 the situation does not change in the case of a wave packet (that is, for a modulated carrier).
 Also in this case, because of the  absence of
 dispersion, all components at different frequencies do not suffer deformation and
propagate with the same  velocity.
\bibitem{note2}
 To be precise, the amplitude of the field is not constant over a
 wavefront. We will neglect such a variation, which limits the
 applicability of our results to a small value of $\theta_0$ and $\rho$.
 This position is justified also because we have to avoid the
 zeros of the field. In fact, when the field is zero, we lose
 information on the direction of propagation, and the Poynting
 vector may have some anomaly.
\end{thebibliography}
\end{document}